\documentclass[twocolumn,amssymb,nobibnotes aps,prl,reprint,superscriptaddress]{revtex4-2}

\usepackage{multirow}
\usepackage{amssymb}
\usepackage{amsmath}
\usepackage{graphicx}
\usepackage{dcolumn}
\usepackage{bm}
\usepackage{wasysym}
\usepackage[mathlines]{lineno}

\begin{document}
\preprint{APS/123-QED}

\title{Testing Local Lorentz Invariance with Laser Tracking of the LAGEOS and LAGEOS II Satellites}

\author{David Lucchesi}
\email[]{david.lucchesi@inaf.it}
\affiliation{Istituto Nazionale di Astrofisica (INAF),  Istituto di Astrofisica e Planetologia Spaziali (IAPS), Via del Fosso del Cavaliere, 100, 00133 Roma, Italy}
\affiliation{Istituto Nazionale di Fisica Nucleare (INFN), Sezione di Tor Vergata, Via della Ricerca Scientifica 1, 00133  Roma, Italy}
\affiliation{Consiglio Nazionale delle Ricerche (CNR), Istituto di Scienza e Tecnologie della Informazione (ISTI),  via G. Moruzzi 1, 56124 Pisa, Italy}
\author{Massimo Visco}
\affiliation{Istituto Nazionale di Astrofisica (INAF),  Istituto di Astrofisica e Planetologia Spaziali (IAPS), Via del Fosso del Cavaliere, 100, 00133 Roma, Italy}
\affiliation{Istituto Nazionale di Fisica Nucleare (INFN), Sezione di Tor Vergata, Via della Ricerca Scientifica 1, 00133  Roma, Italy}
\author{Roberto Peron}
\affiliation{Istituto Nazionale di Astrofisica (INAF),  Istituto di Astrofisica e Planetologia Spaziali (IAPS), Via del Fosso del Cavaliere, 100, 00133 Roma, Italy}
\affiliation{Istituto Nazionale di Fisica Nucleare (INFN), Sezione di Tor Vergata, Via della Ricerca Scientifica 1, 00133  Roma, Italy}
\author{José C. Rodriguez}
\affiliation{Red de Infraestructuras Geod\'esicas, Instituto Geogr\'afico Nacional, Madrid, Spain}
\author{Massimo Bassan}
\affiliation{Dipartimento di Fisica,  Universit\`a di Tor Vergata, Via della Ricerca Scientifica 1, 00133  Roma, Italy}
\affiliation{Istituto Nazionale di Fisica Nucleare (INFN), Sezione di Tor Vergata, Via della Ricerca Scientifica 1, 00133 Roma, Italy}
\author{Giuseppe Pucacco}
\affiliation{Dipartimento di Fisica,  Universit\`a di Tor Vergata, Via della Ricerca Scientifica 1, 00133  Roma, Italy}
\affiliation{Istituto Nazionale di Fisica Nucleare (INFN), Sezione di Tor Vergata, Via della Ricerca Scientifica 1, 00133 Roma, Italy}
\author{Luciano Anselmo}
\affiliation{Consiglio Nazionale delle Ricerche (CNR), Istituto di Scienza e Tecnologie della Informazione (ISTI),  via G. Moruzzi 1, 56124 Pisa, Italy}
\author{Graham Appleby}
\affiliation{British Geological Survey (Hon Research Associate), United Kingdom}
\author{Marco Cinelli}
\affiliation{Istituto Nazionale di Astrofisica (INAF),  Istituto di Astrofisica e Planetologia Spaziali (IAPS), Via del Fosso del Cavaliere, 100, 00133 Roma, Italy}
\affiliation{Istituto Nazionale di Fisica Nucleare (INFN), Sezione di Tor Vergata, Via della Ricerca Scientifica 1, 00133  Roma, Italy}
\author{Alessandro Di Marco}
\affiliation{Istituto Nazionale di Astrofisica (INAF),  Istituto di Astrofisica e Planetologia Spaziali (IAPS), Via del Fosso del Cavaliere, 100, 00133 Roma, Italy}
\affiliation{Istituto Nazionale di Fisica Nucleare (INFN), Sezione di Tor Vergata, Via della Ricerca Scientifica 1, 00133  Roma, Italy}
\author{Marco Lucente}
\affiliation{Istituto Nazionale di Astrofisica (INAF),  Istituto di Astrofisica e Planetologia Spaziali (IAPS), Via del Fosso del Cavaliere, 100, 00133 Roma, Italy}
\affiliation{Istituto Nazionale di Fisica Nucleare (INFN), Sezione di Tor Vergata, Via della Ricerca Scientifica 1, 00133  Roma, Italy}
\author{Carmelo Magnafico}
\affiliation{Istituto Nazionale di Astrofisica (INAF),  Istituto di Astrofisica e Planetologia Spaziali (IAPS), Via del Fosso del Cavaliere, 100, 00133 Roma, Italy}
\affiliation{Istituto Nazionale di Fisica Nucleare (INFN), Sezione di Tor Vergata, Via della Ricerca Scientifica 1, 00133  Roma, Italy}
\author{Carmen Pardini}
\affiliation{Consiglio Nazionale delle Ricerche (CNR), Istituto di Scienza e Tecnologie della Informazione (ISTI),  via G. Moruzzi 1, 56124 Pisa, Italy}
\author{Feliciana Sapio}
\affiliation{Istituto Nazionale di Astrofisica (INAF),  Istituto di Astrofisica e Planetologia Spaziali (IAPS), Via del Fosso del Cavaliere, 100, 00133 Roma, Italy}
\affiliation{Istituto Nazionale di Fisica Nucleare (INFN), Sezione di Tor Vergata, Via della Ricerca Scientifica 1, 00133  Roma, Italy}
\collaboration{SaToR-G Collaboration}
\date{\today}

\begin{abstract}
Violations of Lorentz Invariance, a cornerstone of modern physics, are predicted by theories of quantum gravity and by extensions of General Relativity involving new vector or tensor fields. In the weak-field limit, such a violation would primarily manifest as a non-zero value for the post-Newtonian parameter $\alpha_1$, which is identically zero in General Relativity. We present a new test of Local Lorentz Invariance by searching for this signature in the orbits of the LAGEOS and LAGEOS II satellites. By applying a Phase Sensitive Detection technique to the mean argument of latitude, derived from about 30 years of Satellite Laser Ranging data, we isolate the periodic signal potentially induced by a preferred reference frame aligned with the Cosmic Microwave Background. {Our analysis yields a new constraint $|\alpha_1| \sim 2 \times 10^{-5}$. This result improves upon the previous best limit from Lunar Laser Ranging and provides the most stringent constraint to date on preferred-frame effects in Earth's gravity.}
\end{abstract}

\maketitle

\textit{Introduction}---\label{sec:intro}
Local Lorentz Invariance (LLI) states that the outcome of any local (in space and in time) non-gravitational experiment is independent of the velocity of the freely falling apparatus in which the experiment is carried out. LLI represents a cornerstone in Einstein theory of gravitation, General Relativity (GR) \cite{1916AnP...354..769E} and, in particular, it represents one of the pillars of Einstein Equivalence Principle (EEP) valid in GR and in all metric theories of gravity \cite{2018tegp.book.....W}.
In GR, the only field that mediates the long-range gravitational interaction is the metric tensor \(g_{\mu\nu}\). 
However, in the context of modern unification theories, other fields --- scalar, vector or tensor in their essence --- may come to play a role in mediating the gravitational interaction in addition to the metric tensor of GR.
In these unification theories, the action of a (four-) vector field \(K^{\mu}\) or of other tensor fields \(B_{\mu\nu}\), besides \(g_{\mu\nu}\), is such that the distribution of matter in the universe typically selects a preferred rest frame for local gravitational physics and, consequently, implies a violation of LLI. Conversely, if only one or more scalar fields are present, besides the metric tensor \(g_{\mu\nu}\), as in the case of tensor-multi-scalar theories of gravitation \cite{1992CQGra...9.2093D}, no violation of Lorentz Invariance is expected.
In the framework of the parameterized post-Newtonian (PPN) formalism \cite{1968PhRv..169.1017N,1971ApJ...163..611W,1972ApJ...177..775N,1972ApJ...177..757W}, valid in the Weak-Field and Slow-Motion (WFSM) limit of GR, the Preferred Frame Effects (PFE) are described by the parameters \(\alpha_1\), \(\alpha_2\) and \(\alpha_3\), that are all equal to zero in GR and in tensor-scalar theories of gravity.
LLI and, consequently, PFE, are well tested in the context of high-energy physics experiments \cite{2005LRR.....8....5M} --- see also \cite{2013CQGra..30m3001L} for a review of Lorentz Invariance tests in effective field theories --- but are much more difficult to test in the context of gravitation. 
Current best limits on LLI obtained in the case of Solar System tests, i.e. in the WFSM limit of GR, were achieved \cite{2008ASSL..349..457M} through Lunar Laser Ranging (LLR) \cite{1973Sci...182..229B}: \(\alpha_1=(-7\pm9)\times10^{-5}\) and  \(\alpha_2=(+1.8\pm2.5)\times10^{-5}\); and exploiting the close alignment between the spin of the Sun and the total angular momentum of the Solar system \cite{1987ApJ...320..871N}: \(\mid{\alpha_2}\mid \lesssim 2.4 \times10^{-7}\).
The above limits were obtained by assuming the existence of PFE with respect to the isotropic cosmic microwave background (CMB).
In this paper, under the same assumption, we provide a new constraint on the PPN parameter \(\alpha_1\) from the analysis of the orbital residuals of the geodetic satellites LAGEOS and LAGEOS II over a time interval of approximately 30 years. These are passive satellites very well tracked through the Satellite Laser Ranging (SLR) technique \cite{2002AdSpR..30..135P,2019JGeod..93.2161P}. In particular, we consider the possible existence of PFE due to the motion of the Earth-Sun-satellite system with respect to the CMB radiation and analyze the effect on the time evolution of a combination of Keplerian orbital parameters.
To our knowledge, this is the first measurement that constrains the PPN parameter \(\alpha_1\) using artificial satellite orbits around the Earth.       
In an extended companion paper \cite{2025PhRvD..98d4034L}, details of the analyses performed to obtain the measurement are provided, along with an in-depth analysis of the main sources of systematic error and theoretical implications.

\textit{Orbital effects of PFE}---\label{sec:PFE}
A seminal paper of 1994 \cite{1994PhRvD..49.1693D} suggested that the orbits of some artificial satellites have the potential to improve the upper limit on the \(\alpha_1\) parameter --- down to the \(\sim10^{-6}\) level --- thanks to the appearance of small divisors which enhance the corresponding PFE.
This possibility of reaching a stringent constraint in the PPN parameter linked to the existence of possible PFE is mainly due to the precision of the SLR tracking technique, that allows to reconstruct the orbit of some geodetic satellites at the cm level. 
This is the case for the two LAGEOS satellites \cite{1976anah.iafcR....J,LG1phB,1985JGR....90.9215C,LG-TN-AI-037-89}.
For these reasons their orbital reconstruction \footnote{Of course, it is necessary to develop a dynamic model for the orbit of these satellites that is equally reliable and accurate \cite{Lucchesietal2015,2016AdSpR..57.1928V,2018PhRvD..98d4034V,2019Univ....5..141L}.} has been exploited for fundamental physics measurements over three decades \cite{1996NCimA.109..575C,2004Natur.431..958C,2004cosp...35..232L,2010PhRvL.105w1103L,2014PhRvD..89h2002L,Lucchesietal2015,2019Univ....5..141L,2019arXiv191001941L,2019EPJC...79..872C,2020Univ....6..139L,2021Univ....7..192L}.
Starting from the Lagrangian \(\mathcal{L}\) of \(\mathcal{N}\) ideal proof-masses gravitationally interacting, it can be shown  that the dependency of \(\mathcal{L}\) on the two PPN parameters \(\alpha_1\) and \(\alpha_2\) (if different from zero) will provide non-boost invariant terms depending on the velocities \(\mathbf{v}^0_a\) of the proof masses with respect to some gravitationally preferred rest frame \cite{1994PhRvD..49.1693D}.
The main effects produced if \(\alpha_1\ne0\) are on the eccentricity vector \(\mathbf{e}\) of the satellite's orbit \footnote{This is the Laplace-Runge-Lenz vector that identifies the satellite pericenter direction: a constant of motion in the ideal case of the 2-body Newtonian problem.} and on its mean argument of latitude \(\ell_0=\omega+M\), with \(\omega\) the argument of pericenter and \(M\) the mean anomaly. 
In this work we focus on the effect of LLI violation on $\ell_0$, that is a periodic effect with annual periodicity.
From now on we will refer to \(\ell_0\) simply as the longitude of the satellite or observable.
If LLI is violated, the Lagrangian \(\mathcal{L}\) of the Earth-satellite system contains a perturbing term with the following explicit expression \cite{1994PhRvD..49.1693D}:
\begin{equation}\label{eq:Lag2}
\mathcal{L}_{\alpha_1}=-\frac{{\alpha_1}}{2c^2} \frac{G_Nm_{\oplus}m_S}{r_{\oplus S}}(\mathbf{v}_{\oplus}+\mathbf{w})\cdot (\mathbf{v}_S+\mathbf{v}_{\oplus}+\mathbf{w}),
\end{equation}
where \(m_{\oplus}\) and \(m_S\) are the mass of the Earth and of the satellite, while \(G_N\), \(c\) and \(r_{\oplus S}\) are, respectively, the Newtonian Gravitational constant, the speed of light and the distances between the two masses.
In Eq.  (\ref{eq:Lag2}) \(\mathbf{v}_{\oplus}\) is the velocity of the Earth with respect to the Sun, \(\mathbf{w}\) is the ``absolute" (i.e. with respect to the preferred frame) 
velocity of the Sun, and \(\mathbf{v}_S\) is the orbital velocity of the satellite around the Earth. 
So,  \(\mathbf{v}_{\oplus}+\mathbf{w}\) represents the ``absolute" velocity of the Earth
 while \(\mathbf{v}_S+\mathbf{v}_{\oplus}+\mathbf{w}\) represents the ``absolute" velocity of the satellite.
The following two equations are the Lagrange's perturbation equations in the argument of pericenter and in the mean anomaly:
\begin{equation}
\frac{d\omega}{dt} = -\frac{\cos i}{na^2(1-e^2)^{1/2}\sin i}\frac{\partial\mathcal{R}}{\partial i} +\frac{(1-e^2)^{1/2}}{na^2e} \frac{\partial\mathcal{R}}{\partial e},\label{eq:L_peri}
\end{equation}
\begin{equation}
\frac{d{M}}{dt} = n -\frac{1-e^2}{na^2e}\frac{\partial\mathcal{R}}{\partial e} -\frac{2}{na}\frac{\partial\mathcal{R}}{\partial a},\label{eq:L_mean}
\end{equation}
where \(\mathcal{R}\) is the disturbing function, which is obtained from the perturbing potential of the Lagrangian \footnote{To compute the variation of the orbital elements, a linear perturbation approach is sufficient in which the right-hand side of the above perturbing equations  are evaluated by keeping constant, to their (mean) nominal values, the semi-major axis \(a\), the eccentricity \(e\), the inclination \(i\) and the mean motion \(n=\sqrt{G_Nm_{\oplus}/a^3}\).}.
If we now add up the two rates to construct the observable \(\dot{\ell}_0=\dot{\omega}+\dot{M}\), the terms linked to \(\partial\mathcal{R}/\partial e\) tend to cancel out, and this is truer the smaller the eccentricity of the orbit.
The advantage of considering the sum of the two observables of Eqs. (\ref{eq:L_peri}) and (\ref{eq:L_mean}), is that most of the perturbative effects of the main gravitational and non-gravitational perturbations tend to be reduced on the final observable \(\dot{\ell}_0\) \footnote{This is also true in the case of some relativistic effects, such as Schwarzschild precession \cite{2019agr..book.....S}.}.
By working out the dot product of Eq. (\ref{eq:Lag2}), the disturbing function reduces to:
\begin{equation}
<\mathcal{R}>_{2\pi}=-\frac{\alpha_1}{c^2}\frac{G_Nm_{\oplus}}{a}(\mathbf{w}\cdot\mathbf{v}_{\oplus})\label{eq:disturbing},
\end{equation}
where the notation \(<\mathcal{R}>_{2\pi}\) implies that we take the average over the unperturbed 2-body Keplerian orbit of the satellite around the Earth.
We now insert this last expression into Eqs.  (\ref{eq:L_peri}) and (\ref{eq:L_mean}), and only retain terms that contain \(\frac{\partial\mathcal{R}}{\partial a}\) that are relevant for the effect we seek. 
Using Kepler third law we finally obtain:
\begin{equation}
<\dot{\ell}_0>_{2\pi}=<\dot{\ell}_0>_{2\pi}^{per}-2\alpha_1n\frac{(\mathbf{w}\cdot\mathbf{v}_{\oplus})}{c^2}+ \mathcal{O}(e\alpha_1),\label{eq:long}
\end{equation}
where the first term takes into account possible long-term periodic perturbative effects of both gravitational and non-gravitational nature.
To compute the term related to the PPN parameter \(\alpha_1\) we introduce the plane of the ecliptic as astronomical reference frame with the Sun at the origin, the direction \(\hat{\mathbf{x}}\) toward the Vernal Equinox \(\aries\) and \(\hat{\mathbf{z}}\) normal to the ecliptic plane.  In this frame, the two velocities in Eq. (\ref{eq:long}) take the form (see Table \ref{tab:CMB}):
\begin{equation}
\mathbf{w} = w\left( \cos \beta_{PF} \cos \lambda_{PF} \hat{\mathbf{x}} +  \cos \beta_{PF} \sin \lambda_{PF} \hat{\mathbf{y}} + \sin \beta_{PF} \hat{\mathbf{z}} \right),\label{eq:WPFE}
\end{equation}
\begin{equation}
\mathbf{v}_{\oplus} = {v}_{\oplus}\left( \sin (\lambda_0 + \dot{\lambda}_{\oplus}t) \hat{\mathbf{x}} - \cos (\lambda_0 + \dot{\lambda}_{\oplus}t) \hat{\mathbf{y}}   \right).\label{eq:vTerra}
\end{equation}
\begin{table}[h!]
\caption{Solar-system velocity \(\mathbf{w}\) with respect to the cosmic microwave background. Coordinates are in the ecliptic reference frame. The estimated errors are \(\pm2\) km/s and \(\pm0^{\circ}.01\). Adapted from  \cite{2018tegp.book.....W}.\label{tab:CMB}}
\centering
\begin{ruledtabular} 
\begin{tabular}{lc}
 & Velocity vector \(\mathbf{w}\)  \\
\hline
Absolute value: \(w\)  & 368 km/s  \\
Latitude: \(\beta_{PF}\) & \(-11^{\circ}.13\)   \\
Longitude: \(\lambda_{PF}\) & \(171^{\circ}.55\) \\
\end{tabular}
\end{ruledtabular}
\end{table}
In Eq. (\ref{eq:WPFE}), the coordinates \((\beta_{PF}, \lambda_{PF})\) represent, respectively, the ecliptic latitude and longitude that identify the direction of the preferred frame (in this analysis, that of CMB radiation). 
In Eq. (\ref{eq:vTerra}), \(\lambda_0\) represents the ecliptic longitude of the Earth around the Sun at a fixed epoch, while \(\dot{\lambda}_{\oplus}\) represents its angular orbital rate. This angular rate can be approximated with the Earth's mean motion \(n_{\oplus}=\sqrt{G_NM_{\odot}/a^3_{\odot\oplus}}\), assuming for simplicity a circular Earth orbit (where \(M_{\odot}\) and \(a_{\odot\oplus}\) represent, respectively, the mass of the Sun and the astronomical unit).
Substituting these expressions for the velocities \(\mathbf{w}\) and \(\mathbf{v}_{\oplus}\) into Eq. (\ref{eq:long}), we finally obtain:
\begin{equation}
<\dot{\ell}_0>_{2\pi}=-2\alpha_1n\frac{wv_{\oplus}}{c^2}\cos \beta_{PF}\sin(\lambda_0 +n_{\oplus}t-\lambda_{PF}) + \dots\label{eq:long2}
\end{equation}
where the dots represent minor contributions and we have temporarily removed the unmodeled or poorly modeled periodic effects.

\begin{figure}[h!]
 \includegraphics[width=0.4\textwidth]{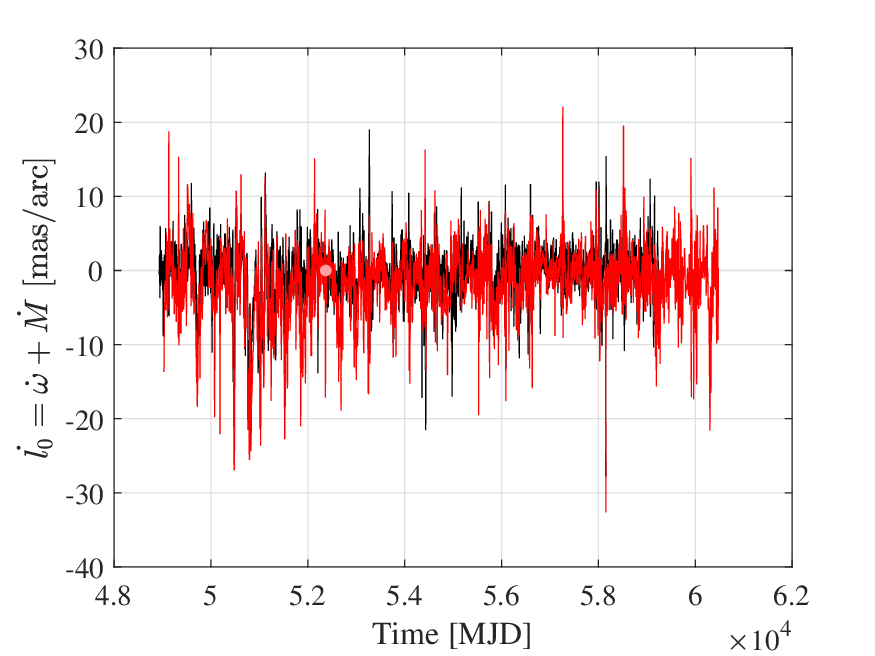}
 \caption{LAGEOS 7-day residuals in the rate of the longitude \({\ell}_0={\omega}+{M}\) versus time (black-GEODYN II, red-SATAN).  \label{fig:residui2}}
 \end{figure}

\begin{figure}[h!]
 \includegraphics[width=0.4\textwidth]{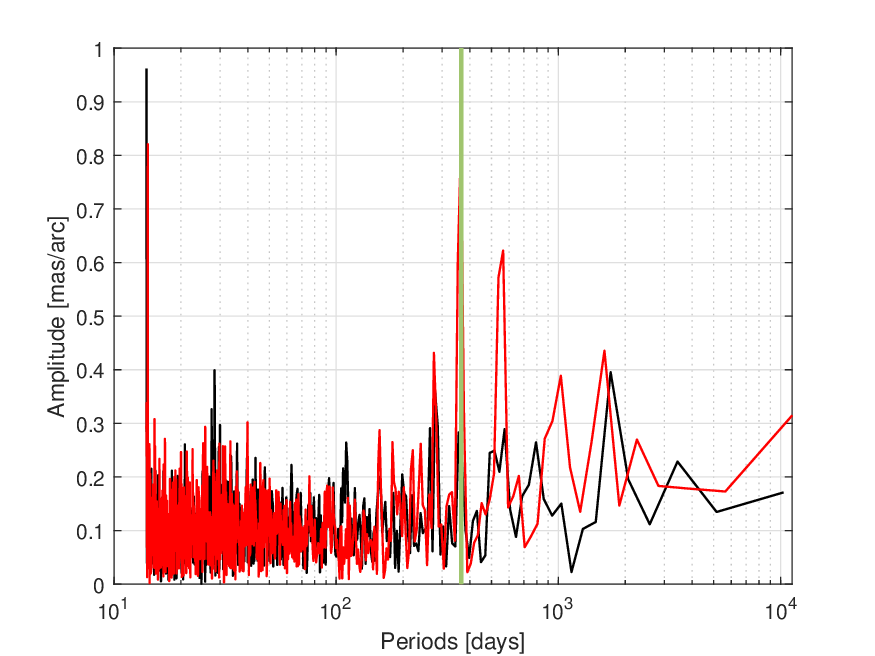}
 \caption{FFT of the residuals in the rate of the longitude \({\ell}_0={\omega}+{M}\) of LAGEOS over the timespan of the analysis. For ease of consultation, the plot is reversed, carrying in abscissa the periods, i.e. 1/frequency. A peak at annual periodicity is clearly visible.
\label{fig:FFT1}}
 \end{figure}

\textit{Orbit Determination and Residuals}.\label{sec:POD}
The Precise Orbit Determination (POD) of the LAGEOS satellites has been made using two independent software, GEODYN II \cite{1998pavlis} and SATAN  \cite{SinclairAppleby1988}, for the data reduction of the satellite tracking data.
The orbit analysis covers a timespan of about 28.3 years, starting from 31 October 1992 (i.e. MJD 48925) in the case of  GEODYN II POD, and about 31.4 years, starting from February 7, 1993 (i.e. MJD 49025) in the case of the POD conducted with SATAN.
Figure \ref{fig:residui2} shows the results, in the case of LAGEOS, for the residuals in our observable \(\dot{\ell}_0=\dot{\omega}+\dot{M}\), introduced to set bounds on the \(\alpha_1\) parameter \footnote{These residuals were obtained by adding the residuals in the rate of the argument of pericenter and in the rate of the mean anomaly of LAGEOS, see Figure 1 in \cite{2025PhRvD..98d4034L}, after removing, on the basis of the rates reported in Table IV of the same paper, the total relativistic precession predicted by GR (not modeled in the POD) on the rates of the argument of pericenter and of the mean anomaly.}.
The black line represents the GEODYN II residuals, while the red line represents the residuals obtained with SATAN.
These residuals represent the variation of the orbital longitude over 7 days: this  is the duration of an “arc”, and is the time unit of these measurements and of our analysis.
The residuals contain the impact of unmodeled perturbing effects on the satellite's orbit \footnote{The overall period of our analysis has been divided into 7-day arcs not causally connected to each other, and the initial state-vector for each arc has been adjusted in the POD procedure to best fit the tracking data}.
The presence of the possible PFE on these residuals is however masked by a plethora of gravitational and non-gravitational effects of both periodic and secular nature.
This is also shown in Figure \ref{fig:FFT1} that represents the frequency domain representation, via the Fast Fourier Transform (FFT), of the residuals of Figure \ref{fig:residui2}. 
{We can take advantage of  the two independent measurements of the longitudes $\ell_0$ of the LAGEOS satellites to reduce the error in the measurement of the parameter \(\alpha_1\). As widely discussed in \cite{2025PhRvD..98d4034L}, the main systematic error on these observable is due to the imperfect knowledge of the Earth's quadrupole coefficient $\bar{C}_{2,0}$  used in the POD.  Following Kozai \cite{1959AJ.....64..367K},  and neglecting the contribution of the terms of  \(\mathcal{O}(e^2)\), we can relate the error $\delta\bar{C}_{2,0}$ with that of the longitude rate $\delta(\dot{\ell}_0)$:}	
\begin{equation}
		\delta(\dot{\ell}_0)|_{\delta\bar{C}_{2,0}} \simeq \frac{3}{2}\sqrt{5}\left(\frac{R_{\oplus}}{a}\right)^2n(3-4\sin^2 i)\delta\bar{C}_{2,0}\label{eq:errC20}.
\end{equation}  
{By using equations (\ref{eq:errC20}) and (\ref{eq:long2}) together, we can treat the longitudes  $\dot{\ell}_0^{LG1}$ and $\dot{\ell}_0^{LG2}$ obtained by the PODs of LAGEOS and LAGEOS II as the left-hand sides in a linear system of two equations in two unknowns.  The two unknowns are the variables $\delta\bar{C}_{2,0}$ and $A_1(t)$. The latter accounts for the possible contribution of $\alpha_1$ at the annual frequency, decoupled from its primary systematic error, $\delta\bar{C}_{2,0}$}. The system to solve is:
\begin{equation}
\left\{\begin{array}{@{}l@{}}
\dot{\ell}_0^{LG1}=K_1^{LG1} \cdot A_1(t) + K_2^{LG1}\cdot \delta\bar{C}_{2,0} \\
\dot{\ell}_0^{LG2}=K_1^{LG2} \cdot A_1(t) + K_2^{LG2}\cdot \delta\bar{C}_{2,0} 
\end{array}\right.\,
\label{equ_comb}
\end{equation}
with $K_1^{LG}$ and $K_2^{LG}$ the coefficients multypling $\alpha_1$ and $\delta\bar{C}_{2,0}$ in Eqs. (\ref{eq:long2}) and (\ref{eq:errC20}).
We are interested in detecting a signal, related to our observable, that varies with annual period and with a well defined phase.
To isolate this signal from the other frequency components, we adopt a Phase-Sensitive Detection (PSD) scheme: a homodyne demodulation at the target frequency (the annual one) and phase, followed by a low-pass filter \footnote{The periodic signal that we wish to detect, after the homodyne detection, splits into a DC component and a periodic component with a frequency double that of the initial one, and both components with an amplitude half of the initial one. This signal is low-pass filtered for the final analysis and measurement.}. In this way we obtain the amplitude of the target signal as a function of time at zero frequency.
In conclusion, applying the phase-sensitive detection to the solution  $A_1(t)$ of the system (\ref{equ_comb}), we finally obtain:
\begin{equation}
\alpha_1 \simeq  <\sin(n_{\oplus}t+\phi_{0}) {A_1(t)}>_{2\pi},\label{eq:LI_long2}
\end{equation}
where \(\phi_0=\lambda_0-\lambda_{PF}\).
This DC component represents the possible violation signal to be extracted from the final data.

\begin{figure}[h!]
 \includegraphics[width=0.45\textwidth]{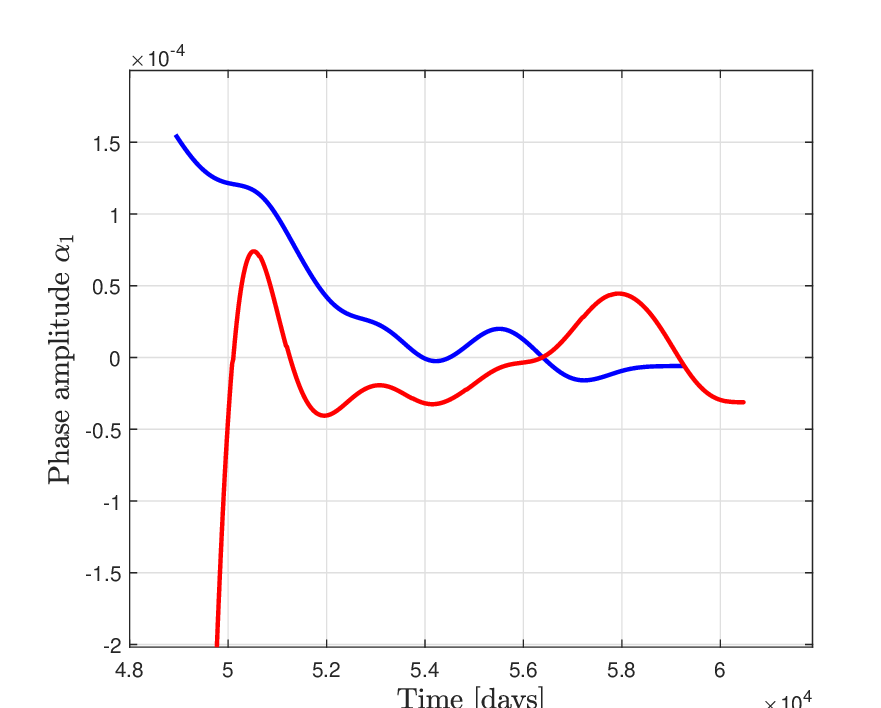}
 \caption{Time behavior of \(A_1(t)\) after the PSD for the in-phase analysis: blue-GEODYN II and red-SATAN. In this specific case we used a 3rd-order low-pass filter, corresponding to an attenuation of about 60 dB per decade of the amplitude of the signal above the cutoff frequency. The low-pass filter removes the higher frequencies and leaves the long-term trend we are looking for. A value of 3000 days was assumed for the integration time of the low-pass filter.
\label{fig:residui3}}
 \end{figure}

\textit{Measurement and Constraint}---\label{sec:measure}
The PSD technique was applied using a reference signal with annual frequency \(f_{0}\simeq 2.738\times 10^{-3}\) days\(^{-1}\) and phase \(\phi_0\). 
The initial phase $\phi_0$ was fixed for each analysis based on the Earth's ecliptic longitude $\lambda_0$ at the respective start dates:  $\lambda_0 \simeq 223^{\circ}.83$ required $\phi_0 = 52^{\circ}.28$ and $\lambda_0 \simeq 318^{\circ}.23$ required $\phi_0 = 146^{\circ}.68$, respectively for GEODYN II and SATAN.
Figure \ref{fig:residui3} shows the results of the in-phase component of \(A_1(t)\). 
The DC component estimated from this plot represents the value of \(\alpha_1\) and, therefore, of a possible violation of the LLI.
Figure \ref{fig:residui3} also shows that PSD filtering has a settling time. To avoid ringing effects, we considered data from day 2000 to day 9000 for each analysis, within which we do not expect a signal altered by the low-pass filter.
From the PSD output we computed the mean and standard deviation of the constrained parameter and obtained:
\begin{equation}
\begin{cases}
\alpha_1 = \left(+2\pm3\right)\times 10^{-5} \text{ with GEODYN II}\label{risultato}\\
\alpha_1 = \left(+1\pm2\right)\times 10^{-5} \text{ with SATAN}.
\end{cases}
\end{equation}

\begin{figure}[h!]
 \includegraphics[width=0.45\textwidth]{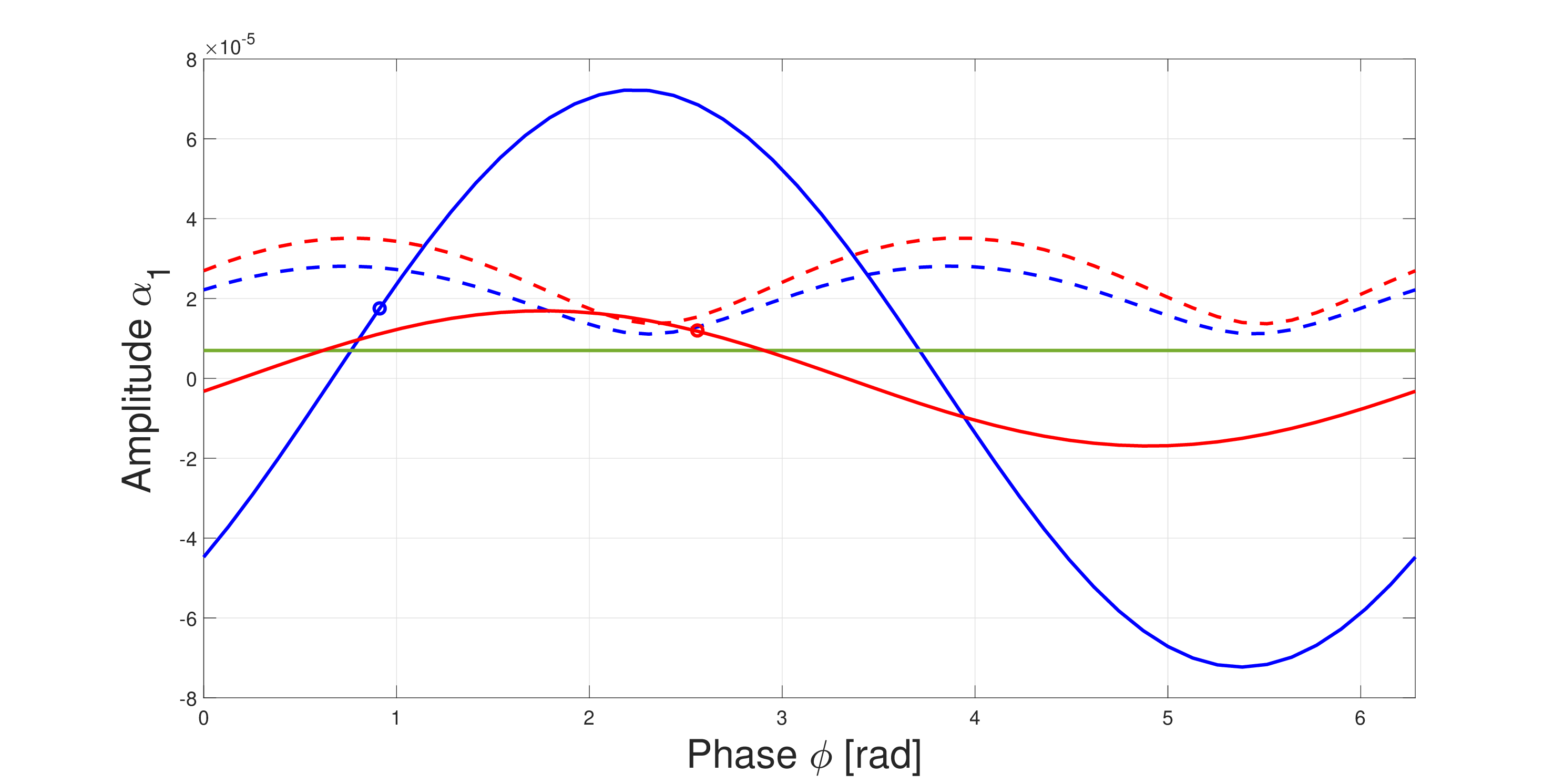}
 \caption{Behavior of \(\alpha_1\) (solid line: blue-GEODYN II and red-SATAN) as the phase of the demodulation sinusoid varies. The frequency of the sinusoid is fixed at the annual value. The blue and red dashed lines represent the corresponding values for the standard deviation. 
The two circles, blue and red, correspond to the values of $\alpha_1$ obtained for GEODYN II and SATAN respectively, see Eq. (\ref{risultato}).
The green continuous line shows the overall estimated systematic error to \(\alpha_1\)  due to both gravitational and non-gravitational perturbations, about \(8\times10^{-6}\).
 The filter parameters are the same as those used in Figure \ref{fig:residui3}.\label{fig:alpha1}}
 \end{figure}

These results are fully compatible with a null value for the PPN parameter \footnote{The constraint proved robust to changes in the filter's order (varied from 3 to 5) and integration time (from 2000 to 5000 days).}.
Figure \ref{fig:alpha1} represents the estimate obtained for the parameter \(\alpha_1\) by further varying the phase of the (ideal) demodulation sinusoid: with
\(\phi=[0, 2\pi]\) while keeping the reference period at 365.25 days.
The solid lines show how the PPN parameter estimate varies in the case of the in-phase analysis while the dashed line shows the variation of the standard deviation in the same time interval of 7000 days.
The values of \(\alpha_1\) at the  phases \(\phi = \phi_0 \simeq 0.91\) rad (blue circle) and \(\phi = \phi_0 \simeq 2.56\) rad (red circle)  correspond to the specific cases of Figure \ref{fig:residui3}.
{Both in the case of the GEODYN II and SATAN analyses, it is worth emphasizing that the maximum values of \(\alpha_1\) are not obtained at the phase \(\phi = \phi_0\), but at a higher value equal to approximately 2.18 rad for GEODYN II, and at a lower value around 1.79 rad for SATAN. This reinforces our conclusion that what we obtain at the expected frequency and phase value for the violation, actually represents a null measurement for the PPN parameter.}
In the figure, the horizontal green line represents the overall systematic error \(\lesssim8\times10^{-6}\) in the measurement of \(\alpha_1\), which we consider to be independent of the phase of the possible violation signal.
{A detailed study of the main systematic errors is reported in the companion paper \cite{2025PhRvD..98d4034L}. The budget is dominated by the uncertainty in the Earth's hexadecapole coefficient $\bar{C}_{4,0}$, which contributes an error of $\sim 6.5\times 10^{-6}$. To assess the impact of non-gravitational perturbations (NGPs), we performed a rigorous end-to-end simulation, processing our high-fidelity NGP models through the entire measurement pipeline. This analysis confirms that the systematic bias from NGPs is negligible, with the largest contribution from SRP model errors being $(-0.4 \pm 4.0) \times 10^{-6}$.}
The detailed error budget described in \cite{2025PhRvD..98d4034L} leads to three significant conclusions: i) it underscores the thoroughness of our non-gravitational force modeling,  ii) our analysis is demonstrably limited by statistics and not by systematics, and iii) the statistical error is due to broadband noise around the annual period.
Hence, the analysis of the systematic errors reinforces the conclusion of a measurement consistent with zero for \(\alpha_1\) \footnote{The estimate of $\alpha_1$ is robust against systematic effects. Varying key parameters, such as those of the dynamical model and the Solar System's velocity relative to the CMB, induces changes down to $\sim10^{-7}$, well below the statistical uncertainty.}.

\textit{Conclusions}---\label{sec:Conclusions}
Local Lorentz Invariance (LLI) is a cornerstone of both the Standard Model of particle physics and General Relativity, and as such, it underpins our fundamental understanding of spacetime. Testing LLI is therefore a powerful probe for new physics. Many alternative theories of gravity introduce additional fields that can mediate the gravitational interaction, potentially violating LLI and giving rise to preferred-frame effects (PFE).
In this Letter, we present a new test of LLI in the gravitational sector by searching for PFE in the orbits of the LAGEOS and LAGEOS II satellites. Our analysis spans nearly three decades of precise Satellite Laser Ranging data and employs two independent orbit determination software for cross-validation. The constraint is derived from a phase-sensitive detection of an annual periodic effect in the orbital residuals of the satellites' mean argument of latitude.
{Using the POD software SATAN, we found \(\alpha_1=(+1\pm2)\times10^{-5}\) for the Parameterized Post-Newtonian (PPN) parameter, while with GEODYN II we obtained a consistent result of \(\alpha_1=(+2\pm3)\times10^{-5}\). 
The total systematic error budget is dominated by the uncertainty in the Earth's hexadecapole harmonic $\bar{C}_{4,0}$ and is estimated to be $\sigma_{\text{sys}} \approx 8 \times 10^{-6}$, which is about 2 times smaller than the statistical uncertainty obtained with SATAN.}
{This result is consistent with a null value for $\alpha_1$, in full agreement with the predictions of General Relativity. The robustness of this null result was further confirmed through extensive testing, detailed in \cite{2025PhRvD..98d4034L}. An analysis of the data using a shorter lock-in integration time (500 days) produced a time series of nearly uncorrelated  $\alpha_1$ estimates that shows no significant temporal variation, confirming the stability of our result. Furthermore, an extreme sensitivity analysis demonstrated that our measurement is insensitive to large, pessimistic variations in key model parameters.}
This work provides the first constraint on the PPN parameter $\alpha_1$ using satellite orbital data in the Earth's gravitational field. It improves upon the previous best constraint from Lunar Laser Ranging \cite{2008ASSL..349..457M} by more than a factor of {four}, placing stringent new limits on the existence of a preferred frame for local gravitational physics and, consequently, on theories that feature additional vector or tensor fields, such as Einstein-aether theory \cite{PhysRevD.70.024003}\footnote{We note that Einstein-aether theory can be regarded as the low-energy limit of Hořava-Lifshitz gravity \cite{2009PhRvL.102p1301H}.}.
In  \cite{2025PhRvD..98d4034L}  we give the bounds on the parameter space and coupling constants of Einstein's-aether theory and a detailed comparison of our result for \(\alpha_1\) with other tests of Lorentz invariance.

\textit{Acknowledgments}---
This work has been performed by the SaToR-G (Satellite Test of Relativistic Gravity) collaboration  funded by the Commissione Scientifica Nazionale (CSN2) of the Istituto Nazionale di Fisica Nucleare (INFN), to which we are very grateful. 
The authors acknowledge the International Laser Ranging Service (ILRS) for providing high-quality laser ranging data of the {LAGEOS and} LAGEOS II satellites {and an anonymous Referee for helpful comments and remarks that have considerably improved the original Letter}.
This work is partially supported by ICSC – Centro Nazionale di Ricerca in High Performance Computing, Big Data and Quantum Computing, funded by European Union – NextGenerationEU.


\begin{thebibliography}{49}%
\makeatletter
\providecommand \@ifxundefined [1]{%
 \@ifx{#1\undefined}
}%
\providecommand \@ifnum [1]{%
 \ifnum #1\expandafter \@firstoftwo
 \else \expandafter \@secondoftwo
 \fi
}%
\providecommand \@ifx [1]{%
 \ifx #1\expandafter \@firstoftwo
 \else \expandafter \@secondoftwo
 \fi
}%
\providecommand \natexlab [1]{#1}%
\providecommand \enquote  [1]{``#1''}%
\providecommand \bibnamefont  [1]{#1}%
\providecommand \bibfnamefont [1]{#1}%
\providecommand \citenamefont [1]{#1}%
\providecommand \href@noop [0]{\@secondoftwo}%
\providecommand \href [0]{\begingroup \@sanitize@url \@href}%
\providecommand \@href[1]{\@@startlink{#1}\@@href}%
\providecommand \@@href[1]{\endgroup#1\@@endlink}%
\providecommand \@sanitize@url [0]{\catcode `\\12\catcode `\$12\catcode
  `\&12\catcode `\#12\catcode `\^12\catcode `\_12\catcode `\%12\relax}%
\providecommand \@@startlink[1]{}%
\providecommand \@@endlink[0]{}%
\providecommand \url  [0]{\begingroup\@sanitize@url \@url }%
\providecommand \@url [1]{\endgroup\@href {#1}{\urlprefix }}%
\providecommand \urlprefix  [0]{URL }%
\providecommand \Eprint [0]{\href }%
\providecommand \doibase [0]{https://doi.org/}%
\providecommand \selectlanguage [0]{\@gobble}%
\providecommand \bibinfo  [0]{\@secondoftwo}%
\providecommand \bibfield  [0]{\@secondoftwo}%
\providecommand \translation [1]{[#1]}%
\providecommand \BibitemOpen [0]{}%
\providecommand \bibitemStop [0]{}%
\providecommand \bibitemNoStop [0]{.\EOS\space}%
\providecommand \EOS [0]{\spacefactor3000\relax}%
\providecommand \BibitemShut  [1]{\csname bibitem#1\endcsname}%
\let\auto@bib@innerbib\@empty
\bibitem [{\citenamefont {{Einstein}}(1916)}]{1916AnP...354..769E}%
  \BibitemOpen
  \bibfield  {author} {\bibinfo {author} {\bibfnamefont {A.}~\bibnamefont
  {{Einstein}}},\ }\bibfield  {title} {\bibinfo {title} {{Die Grundlage der
  allgemeinen Relativit{\"a}tstheorie}},\ }\href
  {https://doi.org/10.1002/andp.19163540702} {\bibfield  {journal} {\bibinfo
  {journal} {Annalen der Physik}\ }\textbf {\bibinfo {volume} {354}},\ \bibinfo
  {pages} {769} (\bibinfo {year} {1916})}\BibitemShut {NoStop}%
\bibitem [{\citenamefont {{Will}}(2018)}]{2018tegp.book.....W}%
  \BibitemOpen
  \bibfield  {author} {\bibinfo {author} {\bibfnamefont {C.~M.}\ \bibnamefont
  {{Will}}},\ }\href@noop {} {\emph {\bibinfo {title} {{Theory and Experiment
  in Gravitational Physics}}}}\ (\bibinfo  {publisher} {{Cambridge University
  Press}},\ \bibinfo {address} {{Cambridge, UK}},\ \bibinfo {year}
  {2018})\BibitemShut {NoStop}%
\bibitem [{\citenamefont {{Damour}}\ and\ \citenamefont
  {{Esposito-Farese}}(1992)}]{1992CQGra...9.2093D}%
  \BibitemOpen
  \bibfield  {author} {\bibinfo {author} {\bibfnamefont {T.}~\bibnamefont
  {{Damour}}}\ and\ \bibinfo {author} {\bibfnamefont {G.}~\bibnamefont
  {{Esposito-Farese}}},\ }\bibfield  {title} {\bibinfo {title}
  {{Tensor-multi-scalar theories of gravitation}},\ }\href
  {https://doi.org/10.1088/0264-9381/9/9/015} {\bibfield  {journal} {\bibinfo
  {journal} {Classical and Quantum Gravity}\ }\textbf {\bibinfo {volume} {9}},\
  \bibinfo {pages} {2093} (\bibinfo {year} {1992})}\BibitemShut {NoStop}%
\bibitem [{\citenamefont {{Nordtvedt}}(1968)}]{1968PhRv..169.1017N}%
  \BibitemOpen
  \bibfield  {author} {\bibinfo {author} {\bibfnamefont {K.}~\bibnamefont
  {{Nordtvedt}}},\ }\bibfield  {title} {\bibinfo {title} {{Equivalence
  Principle for Massive Bodies. II. Theory}},\ }\href
  {https://doi.org/10.1103/PhysRev.169.1017} {\bibfield  {journal} {\bibinfo
  {journal} {Phys. Rev.}\ }\textbf {\bibinfo {volume} {169}},\ \bibinfo {pages}
  {1017} (\bibinfo {year} {1968})}\BibitemShut {NoStop}%
\bibitem [{\citenamefont {{Will}}(1971)}]{1971ApJ...163..611W}%
  \BibitemOpen
  \bibfield  {author} {\bibinfo {author} {\bibfnamefont {C.~M.}\ \bibnamefont
  {{Will}}},\ }\bibfield  {title} {\bibinfo {title} {{Theoretical Frameworks
  for Testing Relativistic Gravity. II. Parametrized Post-Newtonian
  Hydrodynamics, and the Nordtvedt Effect}},\ }\href
  {https://doi.org/10.1086/150804} {\bibfield  {journal} {\bibinfo  {journal}
  {Astrophys. J.}\ }\textbf {\bibinfo {volume} {163}},\ \bibinfo {pages} {611}
  (\bibinfo {year} {1971})}\BibitemShut {NoStop}%
\bibitem [{\citenamefont {{Nordtvedt}}\ and\ \citenamefont
  {{Will}}(1972)}]{1972ApJ...177..775N}%
  \BibitemOpen
  \bibfield  {author} {\bibinfo {author} {\bibfnamefont {J.~K.}\ \bibnamefont
  {{Nordtvedt}}}\ and\ \bibinfo {author} {\bibfnamefont {C.~M.}\ \bibnamefont
  {{Will}}},\ }\bibfield  {title} {\bibinfo {title} {{Conservation Laws and
  Preferred Frames in Relativistic Gravity. II. Experimental Evidence to Rule
  Out Preferred-Frame Theories of Gravity}},\ }\href
  {https://doi.org/10.1086/151755} {\bibfield  {journal} {\bibinfo  {journal}
  {Astrophys. J.}\ }\textbf {\bibinfo {volume} {177}},\ \bibinfo {pages} {775}
  (\bibinfo {year} {1972})}\BibitemShut {NoStop}%
\bibitem [{\citenamefont {{Will}}\ and\ \citenamefont
  {{Nordtvedt}}(1972)}]{1972ApJ...177..757W}%
  \BibitemOpen
  \bibfield  {author} {\bibinfo {author} {\bibfnamefont {C.~M.}\ \bibnamefont
  {{Will}}}\ and\ \bibinfo {author} {\bibfnamefont {J.~K.}\ \bibnamefont
  {{Nordtvedt}}},\ }\bibfield  {title} {\bibinfo {title} {{Conservation Laws
  and Preferred Frames in Relativistic Gravity. I. Preferred-Frame Theories and
  an Extended PPN Formalism}},\ }\href {https://doi.org/10.1086/151754}
  {\bibfield  {journal} {\bibinfo  {journal} {Astrophys. J.}\ }\textbf
  {\bibinfo {volume} {177}},\ \bibinfo {pages} {757} (\bibinfo {year}
  {1972})}\BibitemShut {NoStop}%
\bibitem [{\citenamefont {{Mattingly}}(2005)}]{2005LRR.....8....5M}%
  \BibitemOpen
  \bibfield  {author} {\bibinfo {author} {\bibfnamefont {D.}~\bibnamefont
  {{Mattingly}}},\ }\bibfield  {title} {\bibinfo {title} {{Modern Tests of
  Lorentz Invariance}},\ }\href {https://doi.org/10.12942/lrr-2005-5}
  {\bibfield  {journal} {\bibinfo  {journal} {Living Reviews in Relativity}\
  }\textbf {\bibinfo {volume} {8}},\ \bibinfo {eid} {5} (\bibinfo {year}
  {2005})},\ \Eprint {https://arxiv.org/abs/gr-qc/0502097} {arXiv:gr-qc/0502097
  [gr-qc]} \BibitemShut {NoStop}%
\bibitem [{\citenamefont {{Liberati}}(2013)}]{2013CQGra..30m3001L}%
  \BibitemOpen
  \bibfield  {author} {\bibinfo {author} {\bibfnamefont {S.}~\bibnamefont
  {{Liberati}}},\ }\bibfield  {title} {\bibinfo {title} {{Tests of Lorentz
  invariance: a 2013 update}},\ }\href
  {https://doi.org/10.1088/0264-9381/30/13/133001} {\bibfield  {journal}
  {\bibinfo  {journal} {Classical and Quantum Gravity}\ }\textbf {\bibinfo
  {volume} {30}},\ \bibinfo {eid} {133001} (\bibinfo {year} {2013})},\ \Eprint
  {https://arxiv.org/abs/1304.5795} {arXiv:1304.5795 [gr-qc]} \BibitemShut
  {NoStop}%
\bibitem [{\citenamefont {{M{\"u}ller}}\ \emph {et~al.}(2008)\citenamefont
  {{M{\"u}ller}}, \citenamefont {{Williams}},\ and\ \citenamefont
  {{Turyshev}}}]{2008ASSL..349..457M}%
  \BibitemOpen
  \bibfield  {author} {\bibinfo {author} {\bibfnamefont {J.}~\bibnamefont
  {{M{\"u}ller}}}, \bibinfo {author} {\bibfnamefont {J.~G.}\ \bibnamefont
  {{Williams}}},\ and\ \bibinfo {author} {\bibfnamefont {S.~G.}\ \bibnamefont
  {{Turyshev}}},\ }\bibfield  {title} {\bibinfo {title} {{Lunar Laser Ranging
  Contributions to Relativity and Geodesy}},\ }in\ \href
  {https://doi.org/10.1007/978-3-540-34377-6_21} {\emph {\bibinfo {booktitle}
  {{Lasers, Clocks and Drag-Free Control: Exploration of Relativistic Gravity
  in Space}}}},\ \bibinfo {editor} {edited by\ \bibinfo {editor} {\bibnamefont
  {{H.~Dittus, C.~Lammerzahl, \& S.~G.~Turyshev}}}}\ (\bibinfo {year} {2008})\
  pp.\ \bibinfo {pages} {457--472}\BibitemShut {NoStop}%
\bibitem [{\citenamefont {{Bender}}\ \emph {et~al.}(1973)\citenamefont
  {{Bender}}, \citenamefont {{Currie}}, \citenamefont {{Dicke}}, \citenamefont
  {{Eckhardt}}, \citenamefont {{Faller}}, \citenamefont {{Kaula}},
  \citenamefont {{Mulholland}}, \citenamefont {{Plotkin}}, \citenamefont
  {{Poultney}}, \citenamefont {{Silverberg}}, \citenamefont {{Wilkinson}},
  \citenamefont {{Williams}},\ and\ \citenamefont
  {{Alley}}}]{1973Sci...182..229B}%
  \BibitemOpen
  \bibfield  {author} {\bibinfo {author} {\bibfnamefont {P.~L.}\ \bibnamefont
  {{Bender}}}, \bibinfo {author} {\bibfnamefont {D.~G.}\ \bibnamefont
  {{Currie}}}, \bibinfo {author} {\bibfnamefont {R.~H.}\ \bibnamefont
  {{Dicke}}}, \bibinfo {author} {\bibfnamefont {D.~H.}\ \bibnamefont
  {{Eckhardt}}}, \bibinfo {author} {\bibfnamefont {J.~E.}\ \bibnamefont
  {{Faller}}}, \bibinfo {author} {\bibfnamefont {W.~M.}\ \bibnamefont
  {{Kaula}}}, \bibinfo {author} {\bibfnamefont {J.~D.}\ \bibnamefont
  {{Mulholland}}}, \bibinfo {author} {\bibfnamefont {H.~H.}\ \bibnamefont
  {{Plotkin}}}, \bibinfo {author} {\bibfnamefont {S.~K.}\ \bibnamefont
  {{Poultney}}}, \bibinfo {author} {\bibfnamefont {E.~C.}\ \bibnamefont
  {{Silverberg}}}, \bibinfo {author} {\bibfnamefont {D.~T.}\ \bibnamefont
  {{Wilkinson}}}, \bibinfo {author} {\bibfnamefont {J.~G.}\ \bibnamefont
  {{Williams}}},\ and\ \bibinfo {author} {\bibfnamefont {C.~O.}\ \bibnamefont
  {{Alley}}},\ }\bibfield  {title} {\bibinfo {title} {{The Lunar Laser Ranging
  Experiment}},\ }\href {https://doi.org/10.1126/science.182.4109.229}
  {\bibfield  {journal} {\bibinfo  {journal} {Science}\ }\textbf {\bibinfo
  {volume} {182}},\ \bibinfo {pages} {229} (\bibinfo {year}
  {1973})}\BibitemShut {NoStop}%
\bibitem [{\citenamefont {{Nordtvedt}}(1987)}]{1987ApJ...320..871N}%
  \BibitemOpen
  \bibfield  {author} {\bibinfo {author} {\bibfnamefont {K.}~\bibnamefont
  {{Nordtvedt}}},\ }\bibfield  {title} {\bibinfo {title} {{Probing Gravity to
  the Second Post-Newtonian Order and to One Part in 10 7 Using the Spin Axis
  of the Sun}},\ }\href {https://doi.org/10.1086/165603} {\bibfield  {journal}
  {\bibinfo  {journal} {\apj}\ }\textbf {\bibinfo {volume} {320}},\ \bibinfo
  {pages} {871} (\bibinfo {year} {1987})}\BibitemShut {NoStop}%
\bibitem [{\citenamefont {{Pearlman}}\ \emph {et~al.}(2002)\citenamefont
  {{Pearlman}}, \citenamefont {{Degnan}},\ and\ \citenamefont
  {{Bosworth}}}]{2002AdSpR..30..135P}%
  \BibitemOpen
  \bibfield  {author} {\bibinfo {author} {\bibfnamefont {M.~R.}\ \bibnamefont
  {{Pearlman}}}, \bibinfo {author} {\bibfnamefont {J.~J.}\ \bibnamefont
  {{Degnan}}},\ and\ \bibinfo {author} {\bibfnamefont {J.~M.}\ \bibnamefont
  {{Bosworth}}},\ }\bibfield  {title} {\bibinfo {title} {{The International
  Laser Ranging Service}},\ }\href@noop {} {\bibfield  {journal} {\bibinfo
  {journal} {Adv. Space Res.}\ }\textbf {\bibinfo {volume} {30}},\ \bibinfo
  {pages} {135} (\bibinfo {year} {2002})}\BibitemShut {NoStop}%
\bibitem [{\citenamefont {{Pearlman}}\ \emph {et~al.}(2019)\citenamefont
  {{Pearlman}}, \citenamefont {{Noll}}, \citenamefont {{Pavlis}}, \citenamefont
  {{Lemoine}}, \citenamefont {{Combrink}}, \citenamefont {{Degnan}},
  \citenamefont {{Kirchner}},\ and\ \citenamefont
  {{Schreiber}}}]{2019JGeod..93.2161P}%
  \BibitemOpen
  \bibfield  {author} {\bibinfo {author} {\bibfnamefont {M.~R.}\ \bibnamefont
  {{Pearlman}}}, \bibinfo {author} {\bibfnamefont {C.~E.}\ \bibnamefont
  {{Noll}}}, \bibinfo {author} {\bibfnamefont {E.~C.}\ \bibnamefont
  {{Pavlis}}}, \bibinfo {author} {\bibfnamefont {F.~G.}\ \bibnamefont
  {{Lemoine}}}, \bibinfo {author} {\bibfnamefont {L.}~\bibnamefont
  {{Combrink}}}, \bibinfo {author} {\bibfnamefont {J.~J.}\ \bibnamefont
  {{Degnan}}}, \bibinfo {author} {\bibfnamefont {G.}~\bibnamefont
  {{Kirchner}}},\ and\ \bibinfo {author} {\bibfnamefont {U.}~\bibnamefont
  {{Schreiber}}},\ }\bibfield  {title} {\bibinfo {title} {{The ILRS:
  approaching 20 years and planning for the future}},\ }\href
  {https://doi.org/10.1007/s00190-019-01241-1} {\bibfield  {journal} {\bibinfo
  {journal} {Journal of Geodesy}\ }\textbf {\bibinfo {volume} {93}},\ \bibinfo
  {pages} {2161} (\bibinfo {year} {2019})}\BibitemShut {NoStop}%
\bibitem [{\citenamefont {{Lucchesi}}\ and\ \citenamefont {{et
  al.}}(2026)}]{2025PhRvD..98d4034L}%
  \BibitemOpen
  \bibfield  {author} {\bibinfo {author} {\bibfnamefont {D.~M.}\ \bibnamefont
  {{Lucchesi}}}\ and\ \bibinfo {author} {\bibnamefont {{et al.}}},\ }\bibfield
  {title} {\bibinfo {title} {{A Local Lorentz Invariance test with LAGEOS
  satellites}},\ }\href {https://doi.org/10.1103/PhysRevD.98.044034} {\bibfield
   {journal} {\bibinfo  {journal} {submitted to \prd}\ } (\bibinfo {year}
  {2026})}\BibitemShut {NoStop}%
\bibitem [{\citenamefont {{Damour}}\ and\ \citenamefont
  {{Esposito-Far{\`e}se}}(1994)}]{1994PhRvD..49.1693D}%
  \BibitemOpen
  \bibfield  {author} {\bibinfo {author} {\bibfnamefont {T.}~\bibnamefont
  {{Damour}}}\ and\ \bibinfo {author} {\bibfnamefont {G.}~\bibnamefont
  {{Esposito-Far{\`e}se}}},\ }\bibfield  {title} {\bibinfo {title} {{Testing
  for preferred-frame effects in gravity with artificial Earth satellites}},\
  }\href {https://doi.org/10.1103/PhysRevD.49.1693} {\bibfield  {journal}
  {\bibinfo  {journal} {Phys. Rev. D}\ }\textbf {\bibinfo {volume} {49}},\
  \bibinfo {pages} {1693} (\bibinfo {year} {1994})},\ \Eprint
  {https://arxiv.org/abs/gr-qc/9311034} {gr-qc/9311034} \BibitemShut {NoStop}%
\bibitem [{\citenamefont {{Johnson}}\ \emph {et~al.}(1976)\citenamefont
  {{Johnson}}, \citenamefont {{Lundquist}},\ and\ \citenamefont
  {{Zurasky}}}]{1976anah.iafcR....J}%
  \BibitemOpen
  \bibinfo {editor} {\bibfnamefont {C.~W.}\ \bibnamefont {{Johnson}}}, \bibinfo
  {editor} {\bibfnamefont {C.~A.}\ \bibnamefont {{Lundquist}}},\ and\ \bibinfo
  {editor} {\bibfnamefont {J.~L.}\ \bibnamefont {{Zurasky}}},\ eds.,\
  \href@noop {} {\emph {\bibinfo {title} {{Anaheim International Astronautical
  Federation Congress}}}}\ (\bibinfo {year} {1976})\BibitemShut {NoStop}%
\bibitem [{\citenamefont {{NASA}}(1975)}]{LG1phB}%
  \BibitemOpen
  \bibfield  {author} {\bibinfo {author} {\bibnamefont {{NASA}}},\ }\href@noop
  {} {\emph {\bibinfo {title} {{LAGEOS Phase B Technical Report, NASA Technical
  Memorandum X-64915}}}},\ \bibinfo {type} {Tech. Rep.}\ \bibinfo {number}
  {TMX-64915}\ (\bibinfo  {institution} {Marshall Space Flight Center},\
  \bibinfo {address} {Marshall Space Flight Center, Alabama 35812},\ \bibinfo
  {year} {1975})\ \bibinfo {note} {february 1975}\BibitemShut {NoStop}%
\bibitem [{\citenamefont {{Cohen}}\ and\ \citenamefont
  {{Smith}}(1985)}]{1985JGR....90.9215C}%
  \BibitemOpen
  \bibfield  {author} {\bibinfo {author} {\bibfnamefont {S.~C.}\ \bibnamefont
  {{Cohen}}}\ and\ \bibinfo {author} {\bibfnamefont {D.~E.}\ \bibnamefont
  {{Smith}}},\ }\bibfield  {title} {\bibinfo {title} {{Lageos scientific
  results - Introduction}},\ }\href {https://doi.org/10.1029/JB090iB11p09217}
  {\bibfield  {journal} {\bibinfo  {journal} {J. Geophys. Res.}\ }\textbf
  {\bibinfo {volume} {90}},\ \bibinfo {pages} {9217} (\bibinfo {year}
  {1985})}\BibitemShut {NoStop}%
\bibitem [{\citenamefont {{Fontana}}(1989)}]{LG-TN-AI-037-89}%
  \BibitemOpen
  \bibfield  {author} {\bibinfo {author} {\bibfnamefont {F.}~\bibnamefont
  {{Fontana}}},\ }\href@noop {} {\emph {\bibinfo {title} {{Physical properties
  of LAGEOS II satellite}}}},\ \bibinfo {type} {Tech. Rep.}\ \bibinfo {number}
  {LG-TN-AI-037}\ (\bibinfo  {institution} {Aeritalia},\ \bibinfo {year}
  {1989})\BibitemShut {NoStop}%
\bibitem [{Note1()}]{Note1}%
  \BibitemOpen
  \bibinfo {note} {Of course, it is necessary to develop a dynamic model for
  the orbit of these satellites that is equally reliable and accurate \cite
  {Lucchesietal2015,2016AdSpR..57.1928V,2018PhRvD..98d4034V,2019Univ....5..141L}.}\BibitemShut
  {Stop}%
\bibitem [{\citenamefont {{Ciufolini}}\ \emph {et~al.}(1996)\citenamefont
  {{Ciufolini}}, \citenamefont {{Lucchesi}}, \citenamefont {{Vespe}},\ and\
  \citenamefont {{Mandiello}}}]{1996NCimA.109..575C}%
  \BibitemOpen
  \bibfield  {author} {\bibinfo {author} {\bibfnamefont {I.}~\bibnamefont
  {{Ciufolini}}}, \bibinfo {author} {\bibfnamefont {D.}~\bibnamefont
  {{Lucchesi}}}, \bibinfo {author} {\bibfnamefont {F.}~\bibnamefont
  {{Vespe}}},\ and\ \bibinfo {author} {\bibfnamefont {A.}~\bibnamefont
  {{Mandiello}}},\ }\bibfield  {title} {\bibinfo {title} {{Measurement of
  dragging of inertial frames and gravitomagnetic field using laser-ranged
  satellites.}},\ }\href {https://doi.org/10.1007/BF02731140} {\bibfield
  {journal} {\bibinfo  {journal} {Nuovo Cim. A}\ }\textbf {\bibinfo {volume}
  {109}},\ \bibinfo {pages} {575} (\bibinfo {year} {1996})}\BibitemShut
  {NoStop}%
\bibitem [{\citenamefont {{Ciufolini}}\ and\ \citenamefont
  {{Pavlis}}(2004)}]{2004Natur.431..958C}%
  \BibitemOpen
  \bibfield  {author} {\bibinfo {author} {\bibfnamefont {I.}~\bibnamefont
  {{Ciufolini}}}\ and\ \bibinfo {author} {\bibfnamefont {E.~C.}\ \bibnamefont
  {{Pavlis}}},\ }\bibfield  {title} {\bibinfo {title} {{A confirmation of the
  general relativistic prediction of the Lense-Thirring effect}},\ }\href
  {https://doi.org/10.1038/nature03007} {\bibfield  {journal} {\bibinfo
  {journal} {Nature}\ }\textbf {\bibinfo {volume} {431}},\ \bibinfo {pages}
  {958} (\bibinfo {year} {2004})}\BibitemShut {NoStop}%
\bibitem [{\citenamefont {{Lucchesi}}(2004)}]{2004cosp...35..232L}%
  \BibitemOpen
  \bibfield  {author} {\bibinfo {author} {\bibfnamefont {D.~M.}\ \bibnamefont
  {{Lucchesi}}},\ }\bibfield  {title} {\bibinfo {title} {{The Lense-Thirring
  effect derivation and the LAGEOS satellites orbit analysis with the new
  gravity field solution from CHAMP}},\ }in\ \href@noop {} {\emph {\bibinfo
  {booktitle} {{35th COSPAR Scientific Assembly}}}},\ \bibinfo {series}
  {{COSPAR Meeting}}, Vol.~\bibinfo {volume} {35},\ \bibinfo {editor} {edited
  by\ \bibinfo {editor} {\bibfnamefont {J.-P.}\ \bibnamefont {{Paill{\'e}}}}}\
  (\bibinfo {year} {2004})\ p.\ \bibinfo {pages} {232}\BibitemShut {NoStop}%
\bibitem [{\citenamefont {{Lucchesi}}\ and\ \citenamefont
  {{Peron}}(2010)}]{2010PhRvL.105w1103L}%
  \BibitemOpen
  \bibfield  {author} {\bibinfo {author} {\bibfnamefont {D.~M.}\ \bibnamefont
  {{Lucchesi}}}\ and\ \bibinfo {author} {\bibfnamefont {R.}~\bibnamefont
  {{Peron}}},\ }\bibfield  {title} {\bibinfo {title} {{Accurate Measurement in
  the Field of the Earth of the General-Relativistic Precession of the LAGEOS
  II Pericenter and New Constraints on Non-Newtonian Gravity}},\ }\href
  {https://doi.org/10.1103/PhysRevLett.105.231103} {\bibfield  {journal}
  {\bibinfo  {journal} {Phys. Rev. Lett.}\ }\textbf {\bibinfo {volume} {105}},\
  \bibinfo {pages} {231103} (\bibinfo {year} {2010})}\BibitemShut {NoStop}%
\bibitem [{\citenamefont {{Lucchesi}}\ and\ \citenamefont
  {{Peron}}(2014)}]{2014PhRvD..89h2002L}%
  \BibitemOpen
  \bibfield  {author} {\bibinfo {author} {\bibfnamefont {D.~M.}\ \bibnamefont
  {{Lucchesi}}}\ and\ \bibinfo {author} {\bibfnamefont {R.}~\bibnamefont
  {{Peron}}},\ }\bibfield  {title} {\bibinfo {title} {{LAGEOS II pericenter
  general relativistic precession (1993-2005): Error budget and constraints in
  gravitational physics}},\ }\href {https://doi.org/10.1103/PhysRevD.89.082002}
  {\bibfield  {journal} {\bibinfo  {journal} {Phys. Rev. D}\ }\textbf {\bibinfo
  {volume} {89}},\ \bibinfo {eid} {082002} (\bibinfo {year}
  {2014})}\BibitemShut {NoStop}%
\bibitem [{\citenamefont {{Lucchesi}}\ \emph {et~al.}(2015)\citenamefont
  {{Lucchesi}}, \citenamefont {{Anselmo}}, \citenamefont {{Bassan}},
  \citenamefont {{Pardini}}, \citenamefont {{Peron}}, \citenamefont
  {{Pucacco}},\ and\ \citenamefont {{Visco}}}]{Lucchesietal2015}%
  \BibitemOpen
  \bibfield  {author} {\bibinfo {author} {\bibfnamefont {D.}~\bibnamefont
  {{Lucchesi}}}, \bibinfo {author} {\bibfnamefont {L.}~\bibnamefont
  {{Anselmo}}}, \bibinfo {author} {\bibfnamefont {M.}~\bibnamefont {{Bassan}}},
  \bibinfo {author} {\bibfnamefont {C.}~\bibnamefont {{Pardini}}}, \bibinfo
  {author} {\bibfnamefont {R.}~\bibnamefont {{Peron}}}, \bibinfo {author}
  {\bibfnamefont {G.}~\bibnamefont {{Pucacco}}},\ and\ \bibinfo {author}
  {\bibfnamefont {M.}~\bibnamefont {{Visco}}},\ }\bibfield  {title} {\bibinfo
  {title} {{Testing the gravitational interaction in the field of the Earth via
  satellite laser ranging and the Laser Ranged Satellites Experiment
  (LARASE)}},\ }\href {https://doi.org/10.1088/0264-9381/32/15/155012}
  {\bibfield  {journal} {\bibinfo  {journal} {Class. Quantum Grav.}\ }\textbf
  {\bibinfo {volume} {32}},\ \bibinfo {pages} {155012} (\bibinfo {year}
  {2015})}\BibitemShut {NoStop}%
\bibitem [{\citenamefont {{Lucchesi}}\ \emph
  {et~al.}(2019{\natexlab{a}})\citenamefont {{Lucchesi}}, \citenamefont
  {{Anselmo}}, \citenamefont {{Bassan}}, \citenamefont {{Magnafico}},
  \citenamefont {{Pardini}}, \citenamefont {{Peron}}, \citenamefont
  {{Pucacco}},\ and\ \citenamefont {{Visco}}}]{2019Univ....5..141L}%
  \BibitemOpen
  \bibfield  {author} {\bibinfo {author} {\bibfnamefont {D.~M.}\ \bibnamefont
  {{Lucchesi}}}, \bibinfo {author} {\bibfnamefont {L.}~\bibnamefont
  {{Anselmo}}}, \bibinfo {author} {\bibfnamefont {M.}~\bibnamefont {{Bassan}}},
  \bibinfo {author} {\bibfnamefont {C.}~\bibnamefont {{Magnafico}}}, \bibinfo
  {author} {\bibfnamefont {C.}~\bibnamefont {{Pardini}}}, \bibinfo {author}
  {\bibfnamefont {R.}~\bibnamefont {{Peron}}}, \bibinfo {author} {\bibfnamefont
  {G.}~\bibnamefont {{Pucacco}}},\ and\ \bibinfo {author} {\bibfnamefont
  {M.}~\bibnamefont {{Visco}}},\ }\bibfield  {title} {\bibinfo {title}
  {{General Relativity Measurements in the Field of Earth with Laser-Ranged
  Satellites: State of the Art and Perspectives}},\ }\href
  {https://doi.org/10.3390/universe5060141} {\bibfield  {journal} {\bibinfo
  {journal} {Universe}\ }\textbf {\bibinfo {volume} {5}},\ \bibinfo {pages}
  {141} (\bibinfo {year} {2019}{\natexlab{a}})}\BibitemShut {NoStop}%
\bibitem [{\citenamefont {{Lucchesi}}\ \emph
  {et~al.}(2019{\natexlab{b}})\citenamefont {{Lucchesi}}, \citenamefont
  {{Visco}}, \citenamefont {{Peron}}, \citenamefont {{Bassan}}, \citenamefont
  {{Pucacco}}, \citenamefont {{Pardini}}, \citenamefont {{Anselmo}},\ and\
  \citenamefont {{Magnafico}}}]{2019arXiv191001941L}%
  \BibitemOpen
  \bibfield  {author} {\bibinfo {author} {\bibfnamefont {D.~M.}\ \bibnamefont
  {{Lucchesi}}}, \bibinfo {author} {\bibfnamefont {M.}~\bibnamefont {{Visco}}},
  \bibinfo {author} {\bibfnamefont {R.}~\bibnamefont {{Peron}}}, \bibinfo
  {author} {\bibfnamefont {M.}~\bibnamefont {{Bassan}}}, \bibinfo {author}
  {\bibfnamefont {G.}~\bibnamefont {{Pucacco}}}, \bibinfo {author}
  {\bibfnamefont {C.}~\bibnamefont {{Pardini}}}, \bibinfo {author}
  {\bibfnamefont {L.}~\bibnamefont {{Anselmo}}},\ and\ \bibinfo {author}
  {\bibfnamefont {C.}~\bibnamefont {{Magnafico}}},\ }\bibfield  {title}
  {\bibinfo {title} {{An improved measurement of the Lense-Thirring precession
  on the orbits of laser-ranged satellites with an accuracy approaching the 1\%
  level}},\ }\href@noop {} {\bibfield  {journal} {\bibinfo  {journal} {arXiv
  e-prints}\ ,\ \bibinfo {eid} {arXiv:1910.01941}} (\bibinfo {year}
  {2019}{\natexlab{b}})},\ \Eprint {https://arxiv.org/abs/1910.01941}
  {arXiv:1910.01941 [gr-qc]} \BibitemShut {NoStop}%
\bibitem [{\citenamefont {{Ciufolini}}\ \emph {et~al.}(2019)\citenamefont
  {{Ciufolini}}, \citenamefont {{Paolozzi}}, \citenamefont {{Pavlis}},
  \citenamefont {{Sindoni}}, \citenamefont {{Ries}}, \citenamefont {{Matzner}},
  \citenamefont {{Koenig}}, \citenamefont {{Paris}}, \citenamefont
  {{Gurzadyan}},\ and\ \citenamefont {{Penrose}}}]{2019EPJC...79..872C}%
  \BibitemOpen
  \bibfield  {author} {\bibinfo {author} {\bibfnamefont {I.}~\bibnamefont
  {{Ciufolini}}}, \bibinfo {author} {\bibfnamefont {A.}~\bibnamefont
  {{Paolozzi}}}, \bibinfo {author} {\bibfnamefont {E.~C.}\ \bibnamefont
  {{Pavlis}}}, \bibinfo {author} {\bibfnamefont {G.}~\bibnamefont {{Sindoni}}},
  \bibinfo {author} {\bibfnamefont {J.}~\bibnamefont {{Ries}}}, \bibinfo
  {author} {\bibfnamefont {R.}~\bibnamefont {{Matzner}}}, \bibinfo {author}
  {\bibfnamefont {R.}~\bibnamefont {{Koenig}}}, \bibinfo {author}
  {\bibfnamefont {C.}~\bibnamefont {{Paris}}}, \bibinfo {author} {\bibfnamefont
  {V.}~\bibnamefont {{Gurzadyan}}},\ and\ \bibinfo {author} {\bibfnamefont
  {R.}~\bibnamefont {{Penrose}}},\ }\bibfield  {title} {\bibinfo {title} {{An
  improved test of the general relativistic effect of frame-dragging using the
  LARES and LAGEOS satellites}},\ }\href
  {https://doi.org/10.1140/epjc/s10052-019-7386-z} {\bibfield  {journal}
  {\bibinfo  {journal} {European Physical Journal C}\ }\textbf {\bibinfo
  {volume} {79}},\ \bibinfo {eid} {872} (\bibinfo {year} {2019})},\ \Eprint
  {https://arxiv.org/abs/1910.09908} {arXiv:1910.09908 [gr-qc]} \BibitemShut
  {NoStop}%
\bibitem [{\citenamefont {{Lucchesi}}\ \emph {et~al.}(2020)\citenamefont
  {{Lucchesi}}, \citenamefont {{Visco}}, \citenamefont {{Peron}}, \citenamefont
  {{Bassan}}, \citenamefont {{Pucacco}}, \citenamefont {{Pardini}},
  \citenamefont {{Anselmo}},\ and\ \citenamefont
  {{Magnafico}}}]{2020Univ....6..139L}%
  \BibitemOpen
  \bibfield  {author} {\bibinfo {author} {\bibfnamefont {D.}~\bibnamefont
  {{Lucchesi}}}, \bibinfo {author} {\bibfnamefont {M.}~\bibnamefont {{Visco}}},
  \bibinfo {author} {\bibfnamefont {R.}~\bibnamefont {{Peron}}}, \bibinfo
  {author} {\bibfnamefont {M.}~\bibnamefont {{Bassan}}}, \bibinfo {author}
  {\bibfnamefont {G.}~\bibnamefont {{Pucacco}}}, \bibinfo {author}
  {\bibfnamefont {C.}~\bibnamefont {{Pardini}}}, \bibinfo {author}
  {\bibfnamefont {L.}~\bibnamefont {{Anselmo}}},\ and\ \bibinfo {author}
  {\bibfnamefont {C.}~\bibnamefont {{Magnafico}}},\ }\bibfield  {title}
  {\bibinfo {title} {{A 1\% Measurement of the Gravitomagnetic Field of the
  Earth with Laser-Tracked Satellites}},\ }\href
  {https://doi.org/10.3390/universe6090139} {\bibfield  {journal} {\bibinfo
  {journal} {Universe}\ }\textbf {\bibinfo {volume} {6}},\ \bibinfo {pages}
  {139} (\bibinfo {year} {2020})}\BibitemShut {NoStop}%
\bibitem [{\citenamefont {{Lucchesi}}\ \emph {et~al.}(2021)\citenamefont
  {{Lucchesi}}, \citenamefont {{Anselmo}}, \citenamefont {{Bassan}},
  \citenamefont {{Lucente}}, \citenamefont {{Magnafico}}, \citenamefont
  {{Pardini}}, \citenamefont {{Peron}}, \citenamefont {{Pucacco}},\ and\
  \citenamefont {{Visco}}}]{2021Univ....7..192L}%
  \BibitemOpen
  \bibfield  {author} {\bibinfo {author} {\bibfnamefont {D.}~\bibnamefont
  {{Lucchesi}}}, \bibinfo {author} {\bibfnamefont {L.}~\bibnamefont
  {{Anselmo}}}, \bibinfo {author} {\bibfnamefont {M.}~\bibnamefont {{Bassan}}},
  \bibinfo {author} {\bibfnamefont {M.}~\bibnamefont {{Lucente}}}, \bibinfo
  {author} {\bibfnamefont {C.}~\bibnamefont {{Magnafico}}}, \bibinfo {author}
  {\bibfnamefont {C.}~\bibnamefont {{Pardini}}}, \bibinfo {author}
  {\bibfnamefont {R.}~\bibnamefont {{Peron}}}, \bibinfo {author} {\bibfnamefont
  {G.}~\bibnamefont {{Pucacco}}},\ and\ \bibinfo {author} {\bibfnamefont
  {M.}~\bibnamefont {{Visco}}},\ }\bibfield  {title} {\bibinfo {title}
  {{Testing Gravitational Theories in the Field of the Earth with the SaToR-G
  Experiment}},\ }\href {https://doi.org/10.3390/universe7060192} {\bibfield
  {journal} {\bibinfo  {journal} {Universe}\ }\textbf {\bibinfo {volume} {7}},\
  \bibinfo {eid} {192} (\bibinfo {year} {2021})}\BibitemShut {NoStop}%
\bibitem [{Note2()}]{Note2}%
  \BibitemOpen
  \bibinfo {note} {This is the Laplace-Runge-Lenz vector that identifies the
  satellite pericenter direction: a constant of motion in the ideal case of the
  2-body Newtonian problem.}\BibitemShut {Stop}%
\bibitem [{Note3()}]{Note3}%
  \BibitemOpen
  \bibinfo {note} {To compute the variation of the orbital elements, a linear
  perturbation approach is sufficient in which the right-hand side of the above
  perturbing equations are evaluated by keeping constant, to their (mean)
  nominal values, the semi-major axis \(a\), the eccentricity \(e\), the
  inclination \(i\) and the mean motion \(n=\protect \sqrt {G_Nm_{\oplus
  }/a^3}\).}\BibitemShut {Stop}%
\bibitem [{Note4()}]{Note4}%
  \BibitemOpen
  \bibinfo {note} {This is also true in the case of some relativistic effects,
  such as Schwarzschild precession \cite {2019agr..book.....S}.}\BibitemShut
  {Stop}%
\bibitem [{\citenamefont {{Pavlis}}\ and\ \citenamefont {{et
  al.}}(1998)}]{1998pavlis}%
  \BibitemOpen
  \bibfield  {author} {\bibinfo {author} {\bibfnamefont {D.~E.}\ \bibnamefont
  {{Pavlis}}}\ and\ \bibinfo {author} {\bibnamefont {{et al.}}},\ }\href@noop
  {} {\emph {\bibinfo {title} {{GEODYN II Operations Manual}}}},\ \bibinfo
  {organization} {NASA GSFC} (\bibinfo {year} {1998})\BibitemShut {NoStop}%
\bibitem [{\citenamefont {{Sinclair}}\ and\ \citenamefont
  {{Appleby}}(1988)}]{SinclairAppleby1988}%
  \BibitemOpen
  \bibfield  {author} {\bibinfo {author} {\bibfnamefont {A.~T.}\ \bibnamefont
  {{Sinclair}}}\ and\ \bibinfo {author} {\bibfnamefont {G.~M.}\ \bibnamefont
  {{Appleby}}},\ }\href@noop {} {\emph {\bibinfo {title} {{SATAN: programs for
  the determination and analysis of satellite orbits for SLR data}}}},\
  \bibinfo {type} {Tech. Rep.}\ (\bibinfo {year} {1988})\BibitemShut {NoStop}%
\bibitem [{Note5()}]{Note5}%
  \BibitemOpen
  \bibinfo {note} {These residuals were obtained by adding the residuals in the
  rate of the argument of pericenter and in the rate of the mean anomaly of
  LAGEOS, see Figure 1 in \cite {2025PhRvD..98d4034L}, after removing, on the
  basis of the rates reported in Table IV of the same paper, the total
  relativistic precession predicted by GR (not modeled in the POD) on the rates
  of the argument of pericenter and of the mean anomaly.}\BibitemShut {Stop}%
\bibitem [{Note6()}]{Note6}%
  \BibitemOpen
  \bibinfo {note} {The overall period of our analysis has been divided into
  7-day arcs not causally connected to each other, and the initial state-vector
  for each arc has been adjusted in the POD procedure to best fit the tracking
  data}\BibitemShut {NoStop}%
\bibitem [{\citenamefont {{Kozai}}(1959)}]{1959AJ.....64..367K}%
  \BibitemOpen
  \bibfield  {author} {\bibinfo {author} {\bibfnamefont {Y.}~\bibnamefont
  {{Kozai}}},\ }\bibfield  {title} {\bibinfo {title} {{The motion of a close
  earth satellite}},\ }\href {https://doi.org/10.1086/107957} {\bibfield
  {journal} {\bibinfo  {journal} {Astron. J.}\ }\textbf {\bibinfo {volume}
  {64}},\ \bibinfo {pages} {367} (\bibinfo {year} {1959})}\BibitemShut
  {NoStop}%
\bibitem [{Note7()}]{Note7}%
  \BibitemOpen
  \bibinfo {note} {The periodic signal that we wish to detect, after the
  homodyne detection, splits into a DC component and a periodic component with
  a frequency double that of the initial one, and both components with an
  amplitude half of the initial one. This signal is low-pass filtered for the
  final analysis and measurement.}\BibitemShut {Stop}%
\bibitem [{Note8()}]{Note8}%
  \BibitemOpen
  \bibinfo {note} {The constraint proved robust to changes in the filter's
  order (varied from 3 to 5) and integration time (from 2000 to 5000
  days).}\BibitemShut {Stop}%
\bibitem [{Note9()}]{Note9}%
  \BibitemOpen
  \bibinfo {note} {The estimate of $\alpha _1$ is robust against systematic
  effects. Varying key parameters, such as those of the dynamical model and the
  Solar System's velocity relative to the CMB, induces changes down to $\sim
  10^{-7}$, well below the statistical uncertainty.}\BibitemShut {Stop}%
\bibitem [{\citenamefont {Jacobson}\ and\ \citenamefont
  {Mattingly}(2004)}]{PhysRevD.70.024003}%
  \BibitemOpen
  \bibfield  {author} {\bibinfo {author} {\bibfnamefont {T.}~\bibnamefont
  {Jacobson}}\ and\ \bibinfo {author} {\bibfnamefont {D.}~\bibnamefont
  {Mattingly}},\ }\bibfield  {title} {\bibinfo {title} {Einstein-aether
  waves},\ }\href {https://doi.org/10.1103/PhysRevD.70.024003} {\bibfield
  {journal} {\bibinfo  {journal} {Phys. Rev. D}\ }\textbf {\bibinfo {volume}
  {70}},\ \bibinfo {pages} {024003} (\bibinfo {year} {2004})}\BibitemShut
  {NoStop}%
\bibitem [{Note10()}]{Note10}%
  \BibitemOpen
  \bibinfo {note} {We note that Einstein-aether theory can be regarded as the
  low-energy limit of Hořava-Lifshitz gravity \cite
  {2009PhRvL.102p1301H}.}\BibitemShut {Stop}%
\bibitem [{\citenamefont {{Visco}}\ and\ \citenamefont
  {{Lucchesi}}(2016)}]{2016AdSpR..57.1928V}%
  \BibitemOpen
  \bibfield  {author} {\bibinfo {author} {\bibfnamefont {M.}~\bibnamefont
  {{Visco}}}\ and\ \bibinfo {author} {\bibfnamefont {D.~M.}\ \bibnamefont
  {{Lucchesi}}},\ }\bibfield  {title} {\bibinfo {title} {{Review and critical
  analysis of mass and moments of inertia of the LAGEOS and LAGEOS II
  satellites for the LARASE program}},\ }\href
  {https://doi.org/10.1016/j.asr.2016.02.006} {\bibfield  {journal} {\bibinfo
  {journal} {Advances in Space Research}\ }\textbf {\bibinfo {volume} {57}},\
  \bibinfo {pages} {1928} (\bibinfo {year} {2016})}\BibitemShut {NoStop}%
\bibitem [{\citenamefont {{Visco}}\ and\ \citenamefont
  {{Lucchesi}}(2018)}]{2018PhRvD..98d4034V}%
  \BibitemOpen
  \bibfield  {author} {\bibinfo {author} {\bibfnamefont {M.}~\bibnamefont
  {{Visco}}}\ and\ \bibinfo {author} {\bibfnamefont {D.~M.}\ \bibnamefont
  {{Lucchesi}}},\ }\bibfield  {title} {\bibinfo {title} {{Comprehensive model
  for the spin evolution of the LAGEOS and LARES satellites}},\ }\href
  {https://doi.org/10.1103/PhysRevD.98.044034} {\bibfield  {journal} {\bibinfo
  {journal} {Phys. Rev. D}\ }\textbf {\bibinfo {volume} {98}},\ \bibinfo {eid}
  {044034} (\bibinfo {year} {2018})}\BibitemShut {NoStop}%
\bibitem [{\citenamefont {{Soffel}}\ and\ \citenamefont
  {{Han}}(2019)}]{2019agr..book.....S}%
  \BibitemOpen
  \bibfield  {author} {\bibinfo {author} {\bibfnamefont {M.~H.}\ \bibnamefont
  {{Soffel}}}\ and\ \bibinfo {author} {\bibfnamefont {W.-B.}\ \bibnamefont
  {{Han}}},\ }\href {https://doi.org/10.1007/978-3-030-19673-8} {\emph
  {\bibinfo {title} {{Applied General Relativity}}}}\ (\bibinfo {year}
  {2019})\BibitemShut {NoStop}%
\bibitem [{\citenamefont {{Ho{\v{r}}ava}}(2009)}]{2009PhRvL.102p1301H}%
  \BibitemOpen
  \bibfield  {author} {\bibinfo {author} {\bibfnamefont {P.}~\bibnamefont
  {{Ho{\v{r}}ava}}},\ }\bibfield  {title} {\bibinfo {title} {{Spectral
  Dimension of the Universe in Quantum Gravity at a Lifshitz Point}},\ }\href
  {https://doi.org/10.1103/PhysRevLett.102.161301} {\bibfield  {journal}
  {\bibinfo  {journal} {\prl}\ }\textbf {\bibinfo {volume} {102}},\ \bibinfo
  {eid} {161301} (\bibinfo {year} {2009})},\ \Eprint
  {https://arxiv.org/abs/0902.3657} {arXiv:0902.3657 [hep-th]} \BibitemShut
  {NoStop}%
\end{thebibliography}
%

\end{document}